\documentclass[12pt]{article}

\usepackage[letterpaper, margin=1.0in]{geometry}
\usepackage[utf8]{inputenc}
\usepackage[square, authoryear, numbers]{natbib}
\usepackage{rotating,graphicx}
\usepackage{floatflt}
\usepackage{wrapfig}
\usepackage{amsmath}
\usepackage{color}
\usepackage{caption}
\usepackage{overpic}
\usepackage{subcaption}
\usepackage{lineno}
\usepackage{cite}
\usepackage{hyperref}
\usepackage{setspace}
\usepackage{authblk}
\usepackage{algorithm}
\usepackage{algpseudocode}
\usepackage{setspace}
\usepackage{tabularx}

\newcommand{\MW}{\mathrm{M}_\mathrm{W}}

\title{Kinematic Afterslip Patterns}
\author{Brendan J. Meade}
\affil{Department of Earth \& Planetary Sciences \\ Harvard University \\
    Cambridge, MA USA \\
    meade@fas.harvard.edu}
\date{}

\begin{document}
\maketitle

\abstract{Non-inertial afterslip has been inferred to occur following large earthquakes.  An explanation for this slow slip phenomenon is that coseismically generated stresses induce sliding on parts of a fault surface with velocity-strengthening frictional properties.  Here we develop an alternative explanation for afterslip based on the idea that afterslip may occur on any portion of a fault that exhibits positive residual geometric moment following an earthquake, including sections that ruptured coseismically.  Following a large earthquake, this model exhibits exponential time decay of afterslip and allows for variable sensitivity to coseismic event magnitude and residual geometric moment.  This model provides a partial explanation for the spatial relationship of co- and post-seismic slip associated with the 2011 $\mathrm{M_W}{=}9$ Tohoku-oki earthquake.}



\pagebreak
\section{Introduction}
Geodetic observations have been used to infer the occurrence of slow afterslip following several large earthquakes.  For some events, afterslip has been inferred to be adjacent to regions that slipped coseismically.  For example, the 1992 $\MW{=}7.3$ Landers \citep{shen1994postseismic}, 1999 $\MW{=}7.6$ Izmit \citep{hearn2002dynamics}, and 2001 $\MW{=}7.8$ Kokoxili \citep{wen2012postseismic}.  An explanation for this spatially distinct behavior is that afterslip tends to occur along sections of faults characterized by coseismic stress changes that promote additional slip and that have velocity-strengthing frictional properties \citep[e.g.,][]{rice1983earthquake, rubin2005earthquake}. 

However, the spatial separation of co- and post-seismic slip is not a universal observation.  For example, the 2011 $\MW{=}9.0$ Tohoku-oki \citep{ozawa2011coseismic, johnson2012challenging, diao2014overlapping} 2015 $\MW{=}8.3$ Illapel earthquake \citep{barnhart2016coseismic, shrivastava2016coseismic}, and 2015 $\MW{=}8.6$ Nias \citep{qiu2019coseismic} earthquakes have all been inferred to exhibit afterslip that spatially overlaps with regions of coseismic slip.  In this work, we will neglect any physical or geodetic imaging biases that might potentially distinguish inferred behaviors between continental and subduction zone settings.  This inferred behavior has been non-trivial to explain with traditional rate and state frictional models where slip.  This modeling difficulty was noted particularly clearly by \citet{johnson2012challenging}, who noted that the relationship between co- and post-seismic slip for this event ``is inconsistent with a simple model consisting of velocity-weakening patches obeying simple rate-state friction surrounded by a velocity strengthening interface.''

Here we describe a model for the spatial location and magnitude of time-dependent afterslip based not on rate-and-state friction but rather on the idea that afterslip serves to release accumulated geometric moment.  In this model, the critical parameters for determining the location of afterslip are not the frictional properties of fault rocks (e.g., $a{-}b$ values, coefficient of static friction) but rather the spatial distribution of residual geometric moment that has not been released coseismically.  Geometric moment is a kinematic quantity defined as the product of slip and area and can be used as a descriptive metric of earthquake size and kinematics \citep{amelung1997large, ben2001quantification} without reference to the material properties of faults or bulk rock.  This model formulation allows for afterslip to possibly, but not necessarily, occur in regions that may overlap with coseismic rupture depending on the kinematic history of faulting and geometric moment accumulation consistent with the idea that all parts of a fault may be involved in active slip events \citep{passelegue2020initial}.

\section{A geometric moment based model for time-dependent afterslip}


The foundational idea behind this model is that afterslip may release additional geometric moment that has not been released coseismically.  Over some interval of time, geometric moment is accumulated due to on-fault slip deficits and released by coseismic sense slip events.  At some time, $t$ the total geometric moment is given by $m_i(t)$ given by the difference between the accumulated and release geometric moment, and we define the residual geometric moment as $m_i^\mathrm{r} = m_i(t_{+})$, where $t_{+}$ is the instant in time after an earthquake.

In the model proposed here, afterslip may occur following each earthquake and may release additional accumulated geometric moment in regions where there is an excess of accumulated geometric moment $m_i^\mathrm{r}>0$.  This could be a section of the fault system that did not slip coseismically or one where a coseismic event occurred but did not slip sufficiently to release all previously accumulated geometric moment. 

To determine how quickly afterslip might occur, we assume that a fault patch may experience an enhanced sliding velocity, $v_i$, during some nominally postseismic interval and that the rate of sliding will be fastest immediately following the earthquake, decaying monotonically in time.  Throughout, we assume that for this numerical treatment, the fault surface has been discretized into patches referenced as $i$.  As afterslip occurs, geometric moment is released at a rate, $dm^\mathrm{r}_i / dt = a_i v_i$.  As additional geometric moment is released, there is less available to drive additional afterslip, and we posit the simple idea that the rate change in afterslip velocity may be proportional to the rate change in residual geometric moment,

\begin{equation}
    \frac{dv_i}{dt}=-c^2f(m^0_i)\frac{dm^\mathrm{r}_i}{dt}
\end{equation}

where $c$ is a constant contributing to the rate of afterslip decay and $f(m^\mathrm{r}_i)$ is a function that describes the sensitivity of the sliding rate to the residual geometric moment.  This differential equation has the solution $v_i(t) = v^0_i e^{-c^2 f(m^\mathrm{r}_i) a_i t}$ where, $v^0_i$, is the sliding velocity at $t=0$.  To constrain this initial sliding velocity we assume that the ratio of geometric moment release during afterslip to that released coseismically, $\gamma = m^\mathrm{c}/m^\mathrm{a}$ can be estimated \citep[e.g.,][]{ozawa2011coseismic}.  The first step in using this global constraint to estimate $v^0_i$ is to note that the total released afterslip moment on a single fault element is,

\begin{equation}
    m_i^\mathrm{a} = a_i v^0_i \int_0^\infty e^{-c^2 f(m^\mathrm{r}_i) a_i t} dt  = \frac{v^0_i}{c^2 f(m^\mathrm{r}_i)}.
\end{equation}

Summing over all elements, we can write the total afterslip moment as a function of the coseismic moment release as, $\gamma m^\mathrm{c} = \sum_{i \in n} v^0_i / (c^2f(m^\mathrm{r}_i))$. Now we suppose that the initial afterslip velocity is a function of the coseismic magnitude, $g(m_i^\mathrm{r})$, and some function of the accumulated geometric moment $h(m^\mathrm{c})$ giving us, $v^0_i = \mathcal{V} g(m_i^\mathrm{r}) h(m^\mathrm{c})$.  With this assumption, we can express the leading coefficient in the definition of the initial sliding velocity as,

\begin{equation}
    \mathcal{V} = \frac{c^2 \gamma m^\mathrm{c}}{\sum_{i \in n} g(m_i^\mathrm{r}) h(m^\mathrm{c})/f(m^\mathrm{r}_i)}
\end{equation}

and general form for the afterslip sliding velocity as,

\begin{equation}
    v'_i(t) = \frac{c^2 \gamma m^\mathrm{c}g(m_i^\mathrm{r}) h(m^\mathrm{c})}{\sum_{i \in n} g(m_i^\mathrm{r}) h(m^\mathrm{c})/f(m^\mathrm{r}_i)} e^{-c^2 f(m^\mathrm{r}_i) a_i t}.
    \label{eq:afterslip_velocity}
\end{equation}

%

There are few compact solutions for arbitrary $f(m_i^\mathrm{r}), g(m_i^\mathrm{r})$, and $h(m_i^\mathrm{c})$, although for the $f(m_i^\mathrm{r})=g(m_i^\mathrm{r})$ case, $\mathcal{V} = c^2 \gamma / n$ where $n$ is the number of fault patches involved in the afterslip process.

\section{Applications}
\subsection{A two-dimensional vertical strike-slip fault}
\begin{figure}[ht]
    \centerline{\includegraphics[width=1.00\textwidth]{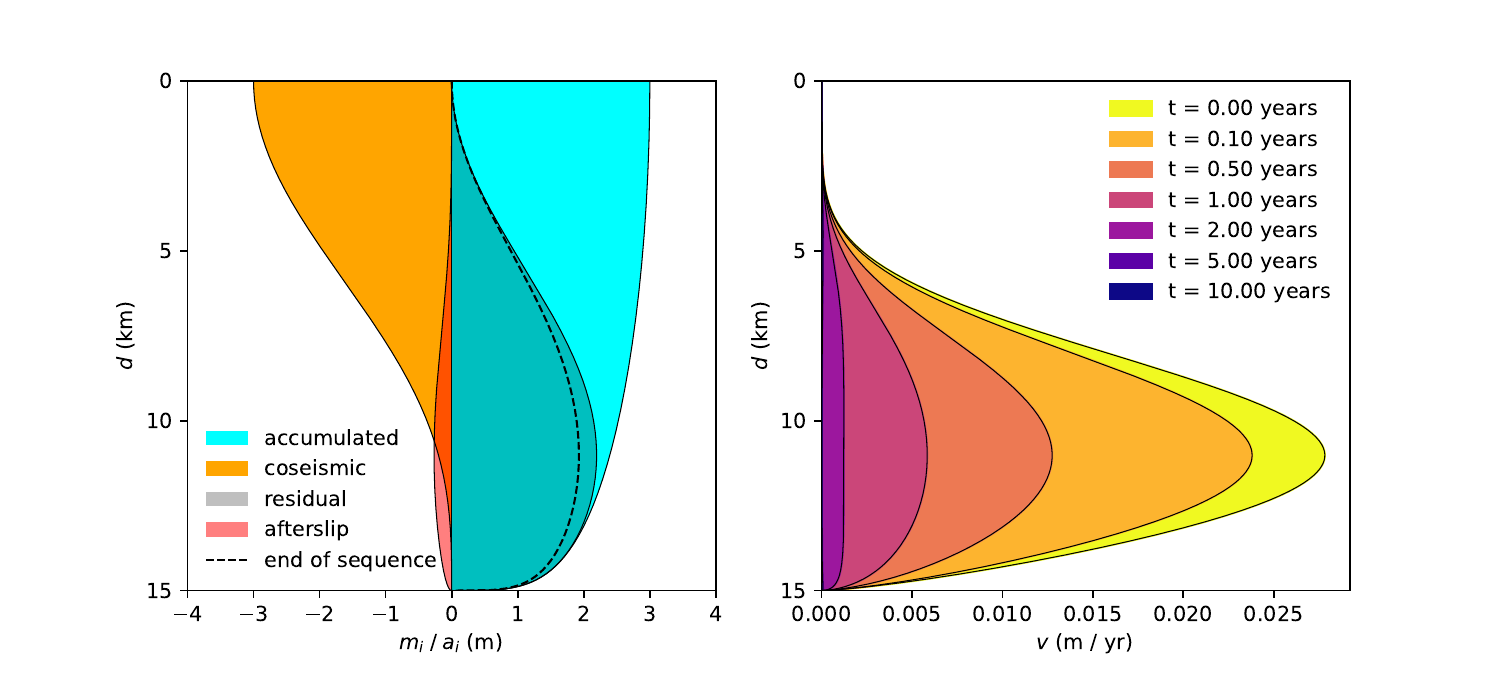}}
    \caption{Afterslip scenario for a two-dimensional vertical strike-slip fault.  The left-side figure shows the geometric moment associated with five stages: 1) positive pre-earthquake accumulated geometric moment (cyan), 2) negative coseismic released geometric moment, 3) residual geometric moment after the coseismic slip (gray), 4) geometric moment release from afterslip (red), and 5) remaining geometric moment after accumulation-coseismic-afterslip sequence (dashed black line).  The right-side figure shows the afterslip velocities at seven times following the coseismic event.  Velocities are fastest immediately after the event, decaying in magnitude in time.  In this example, afterslip occurs at depths of significant and negligible coseismic slip.}
    \label{fig:kinematic_afterslip_m_velocities}
\end{figure}

To illustrate the behavior of the model, we first consider synthetic experiments on a two-dimensional representation of an infinitely long strike-slip fault.  In these experiments, we generate an idealized representation of interseismic slip deficit and coseismic slip release that allows for the calculation of residual geometric moment and the time-dependent afterslip response.  Interseismically accumulated geometric moment is assumed to taper from five meters near the free surface to zero at the assumed base of the fault at 15 km depth (figure \ref{fig:kinematic_afterslip_m_velocities}).  Coseismic geometric moment release is similarly localized at the near-surface, with a maximum of five meters, and decreases with depth at a rate faster than that of the accumulated geometric moment.  These spatial patterns produce a distribution of residual geometric moment that is positive everywhere (except at 0 and 15 km depths) and has a maximum positive value near 12 km depth.  This distribution has little residual geometric moment above 5 km depth, but below 10 km depth, the residual geometric moment is only slightly reduced from the accumulated preseismic geometric moment because there was little coseismic slip in this region.

\begin{figure}[ht]
    \centerline{\includegraphics[width=1.00\textwidth]{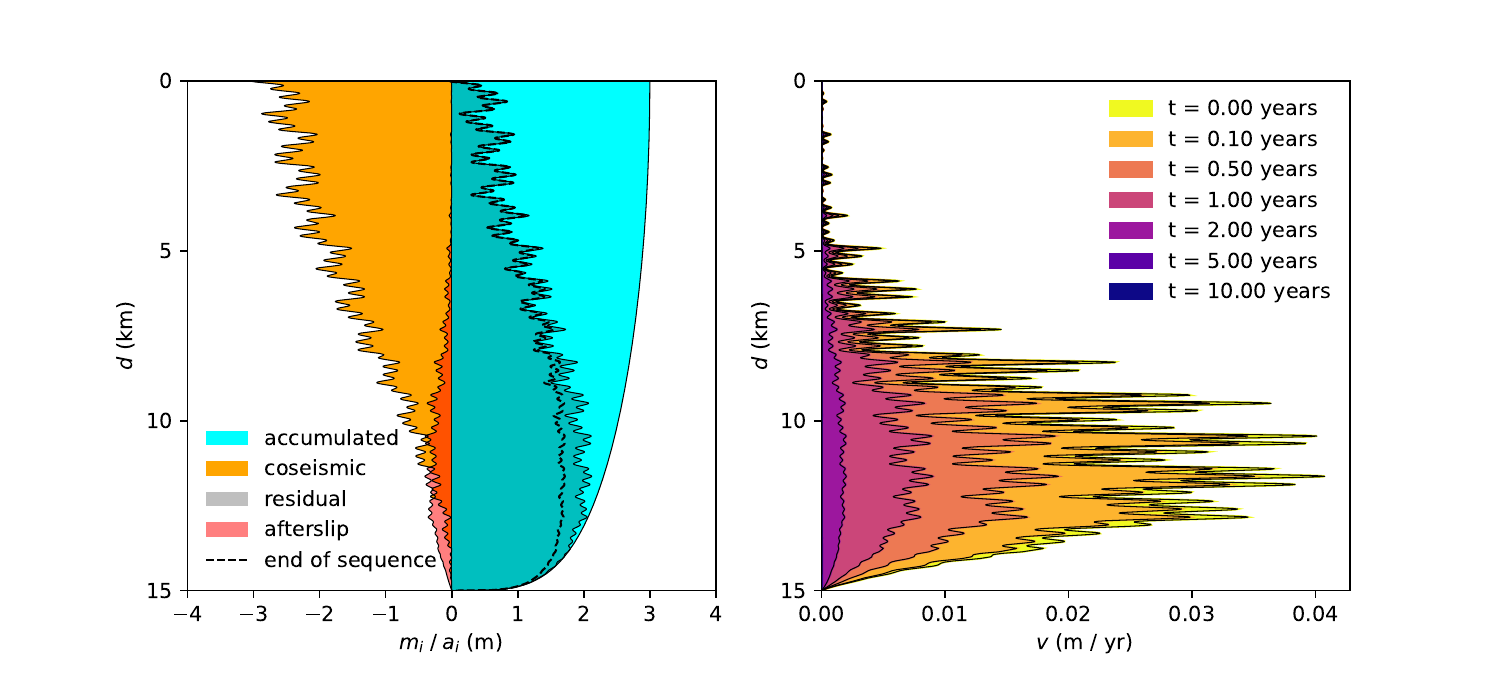}}
    \caption{Same as figure \ref{fig:kinematic_afterslip_m_velocities} except with a spatially rough coseismic slip distribution.  Note that despite positive residual geometric moment at all depths, afterslip is localized at depths ${>}7$ km due to the strong power-law sensitivity of this case.}
    \label{fig:kinematic_afterslip_m_velocities_rough}
\end{figure}

Applying the kinematic afterslip model, we find that immediately following the earthquake, afterslip occurs in all regions of positive residual geometric moment.  Afterslip rates depend on $f(m^\mathrm{r})$ and $g(m^\mathrm{r})$ (equation \ref{eq:afterslip_velocity}), but are fastest in regions with higher residual geometric moment when these functions have a power-law form, $(m^\mathrm{r})^\psi$, where $\psi>1$. Larger values of $\gamma$ increase the initial rate of slip for the regions with the greatest residual geometric moment.  One implication of this behavior is that if the residual geometric moment is spatially rough due to spatially heterogeneous coseismic slip (figure \ref{fig:kinematic_afterslip_m_velocities_rough}), then afterslip will tend to homogenize any roughness in the residual geometric moment distribution because the regions with higher residual geometric moment will slip faster and catch up to more slowly slipping regions that initially have lower residual geometric moment.  Note that in these examples, the afterslip does not release all of the residual geometric moment.  That need not be the case if the coseismic moment release is large enough to release a near majority of the near accumulated geometric moment.

\subsection{The 2011 Tohoku-oki earthquake}
\begin{figure}[ht]
    \centerline{\includegraphics[width=0.80\textwidth]{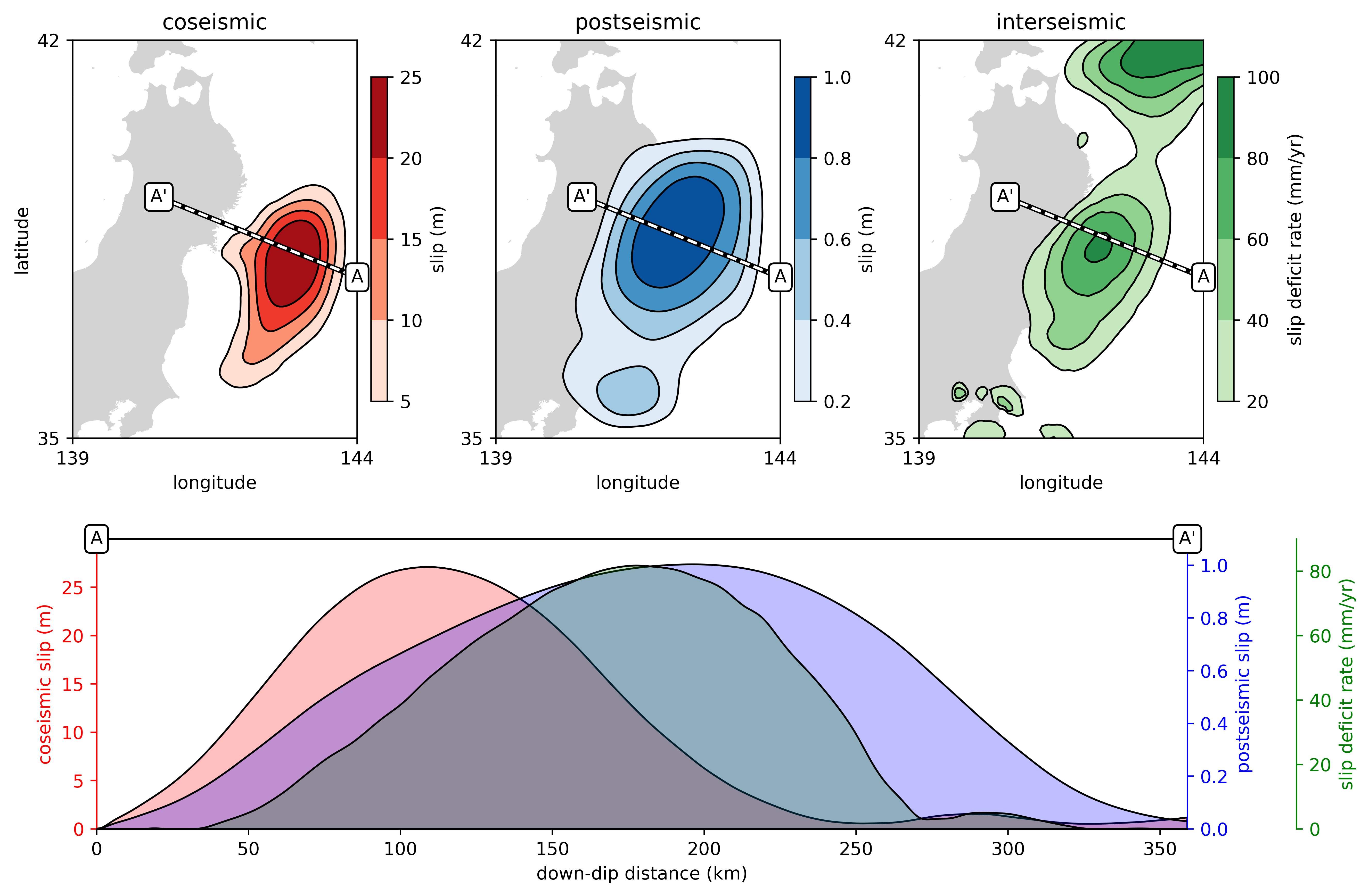}}
    \caption{Published estimates of coseismic slip, postseismic afterslip, and pre-earthquake slip deficit rates along the Japan subduction zone in the vicinity of the 2011 Tohoku-oki earthquake.  Map views of on-fault coseismic and postseismic slip distribution estimates \citep{ozawa2011coseismic} are shown in the upper left and upper central panels, respectively, and interseismic slip deficit rates are shown in the upper right panel \citep{loveless2010geodetic}.  The lower panel shows a profile from all three fields extracted along the A to A' transect, indicated by the white and black dashed line in the upper panel.  Note that the vertical axes are distinct for each of the three quantities, coseismic (red), postseismic (blue), and interseismic (green).  The extracted profiles indicated substantial overlap between coseismic slip and postseismic afterslip at down-dip distances from 50-250 km, with afterslip extending a further 100 km down-dip.  The pre-earthquake interseismic slip deficit spans both the lowermost part of the coseismic rupture region and the central/uppermost part of the afterslip region.}
    \label{fig:japan_sz_distributions}
\end{figure}

The $\MW{=}9$ Tohoku-oki earthquake occurred off the east coast of Japan, where both its coseismic and afterslip motions were observed with the GEONET geodetic network \citep{sagiya2004decade}.  These observations of surface motions have been combined with elastic/viscoelastic models and a range of estimation methods to infer the distribution of both coseismic slip during the fast part of the earthquake afterslip in the weeks \citep[e.g.,][]{ozawa2011coseismic} to months that followed \citep[e.g.,][]{yamagiwa2015afterslip}.  It is these inferred spatial patterns and their relationships to one another that constitute the basis of the application of the kinematic afterslip model to the Tohoku-oki earthquake.  In this context, we note that there is no unique representation of the coseismic and post-earthquake slip distributions for this event.  Indeed, among the large number of different slip inferences, there are significant distinctions, including both large magnitude \citep[e.g.,][]{wei2012sources} and negligible \citep[e.g.,][]{ozawa2011coseismic} coseismic slip at the up-dip limit of the Japan subduction zone.  Differences between these inferences may be due to model choices, data choices, and assumptions that enter into how the inference process is regularized.  Here we make no attempt to parse, nor rationalize, the diversity of published slip estimates.  Instead, we analyze just a single coseismic-afterslip estimate pair that were developed using the same elastic model and inference method for both time intervals \citep{ozawa2011coseismic}.  These slip distributions are notable for two reasons.  First, in this model, coseismic slip distribution does not exhibit significant (${>}5$m) near trench slip (figure \ref{fig:japan_sz_distributions}).  This may be due to the low resolving power of GPS data near the trench \citep{loveless2011spatial} or regularization choices.  For this reason, we do not focus on the near-trench patterns of either coseismic or postseismic slip.  A second notable feature is the spatial relationship between coseismic slip and afterslip (figure \ref{fig:japan_sz_distributions}).  While the inferred coseismic slip reaches an inferred maximum of ${\sim}25$ m at a down-dip distance (this distance is used throughout) of 100 km and the postseismic slip distribution reaches a maximum of ${\sim}1$ m near 175 km, there is qualitatively a significant amount of spatial overlap between the two regions with the area of coseismic slip exceeding 5 meters extending from 30 to 205 km and the region of coseismic slip exceeding 20\% of the maximum value extending from 40 to 310 km.  An interpretation of these slip distributions is that a non-trivial amount of afterslip may have occurred in the region that ruptured coseismically and also along a down-dip extension of the subduction zone interface.  

\begin{figure}[ht]
    \centerline{\includegraphics[width=0.70\textwidth]{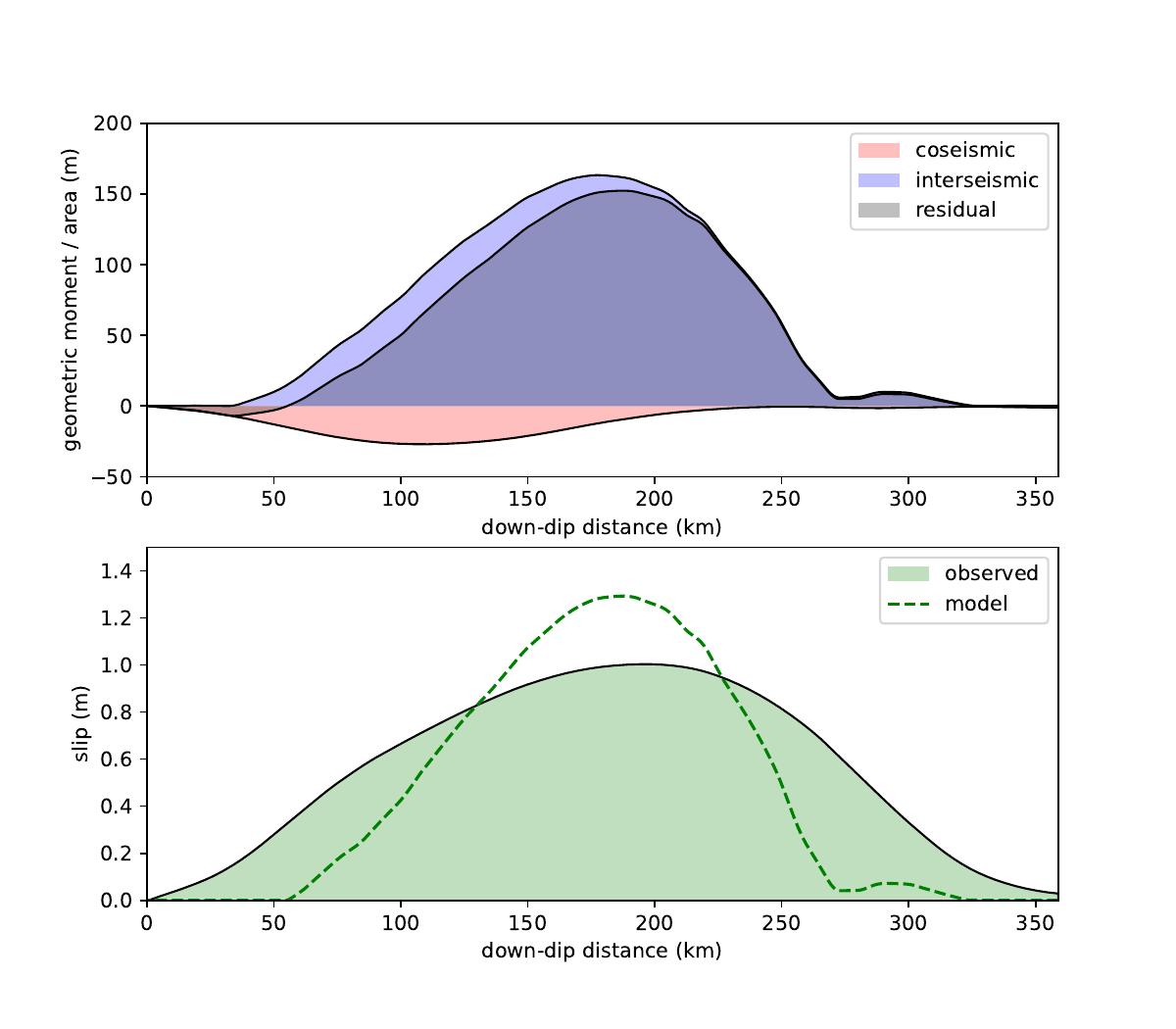}}
    \caption{Upper panel: coseismic moment release (red shaded region), interseismic moment accumulation (blue shaded region), and residual (accumulation - release) geometric moment (gray shaded region).  Lower panel: inferred postseismic slip distribution \citep{ozawa2011coseismic} (green shaded region), kinematic afterslip model prediction of afterslip distribution (green dashed line).}
    \label{fig:japan_after_slip_model}
\end{figure}

We apply the kinematic afterslip model to inferences of fault behavior proximal to the 2011 Tohoku-oki earthquake along a two-dimensional profile representation through the following set of steps.  1) Assume that geometric moment accumulated for 2,000 years prior to the earthquake at a constant rate with a spatial distribution pre-earthquake inferred from block models of pre-earthquake interseismic activity \citep{loveless2010geodetic}.  2) Determine a representation of residual geometric moment after the Tohoku-oki earthquake by subtracting a coseismic slip distribution \citep{ozawa2011coseismic} from the accumulated geometric moment.  3) Calculate postseismic response along a discretized representation of the subduction zone, assuming simple, functional forms of $f(m^\mathrm{r}_i) = g(m^\mathrm{r}_i) = m^\mathrm{r}_i$ and $h(m^\mathrm{c})=m^\mathrm{c}$ for the kinematic afterslip model.  Notably, this simple choice of functional forms for the afterslip model happens to yield spatially uniform cumulative afterslip at long times after the coseismic event.  The predicted distribution of postseismic afterslip (figure \ref{fig:japan_after_slip_model}) is qualitatively similar to the inferred afterslip distribution featuring afterslip both in regions that slipped significantly coseismically and down dip as well as exhibiting a peak afterslip amplitude of 1.3 m at a down-dip distance of 190 km, as compared to the geodetically inferred afterslip distribution with a peak amplitude of 1 m at a down-dip distance of 210 km.  However, the predicted afterslip is more spatially localized compared with the geodetically inferred slip distribution.  The reason for this is that there is not significant enough residual geometric moment at the up- and down-dip extents of the subduction zone to drive afterslip.  If we wanted to assume that this particular set of geodetically constrained inferred coseismic and afterslip distributions were worth fitting in detail, then this discrepancy might be ascribed to an insufficient representation of interseismic geometric moment accumulation in time and/or space. While the purpose of this paper is to demonstrate the first-order explanatory power of the kinematic afterslip model, it is possible that the model could be optimized to not only solve for $g$, $f$, and $h$ but also the spatial distribution of pre-earthquake accumulated geometric moment. 

\section{Discussion}
The two primary goals of the kinematic afterslip model described here are: 1) To develop an explanation of afterslip as a result of releasing accumulated geometric moment rather than stress and 2) to provide a possible explanation for the spatial pattern of afterslip following the Tohoku-oki earthquake.  Stress-driven models of afterslip have been used to model geodetic observations following both continental \citep[e.g.,][]{freed2007afterslip} and subduction zone earthquakes \citep[e.g.,][]{hsu2006frictional}.  In these cases, the rate of afterslip is generally governed by the rate-and-state friction equations \citep[e.g.,][]{marone1991mechanics}, features a logarithmic decay rate of slip in time (table \ref{tab:stress_vs_geometric_moment}), and afterslip is generally predicted to be localized in regions adjacent to that of the coseismic rupture for two primary reasons.  First, the coseismic region generally experiences a shear stress drop due to the earthquake, and the regions adjacent to the main shock may experience an increase in shear stress because of the resulting elastic deformation of the surrounding region. Second, the frictional properties of rocks at the fault interface may vary with depth, with velocity-weakening regions enabling fast coseismic slip and velocity-strengthening regions more likely to host slow slip during the post-earthquake interval.  Taken together, these two features of the rate-and-state friction model tend to localize afterslip in regions adjacent to the coseismic rupture where there has been an increase in shear stress and rocks exhibit velocity strengthening behavior \citep[e.g.,][]{marone1991mechanics}.

In contrast, the kinematic afterslip model does not directly dependent on rock properties and instead depends on geometric moment as a representation of how much slip has and has not occurred along a specific section of a fault.  This means that afterslip may occur anywhere along a fault, not just in regions that have velocity-weakening frictional properties and have experienced increases in stresses that promote sliding.  This is conceptually similar to arguments that have been advanced to explain results from laboratory experiments where stored strain energy has been suggested to play a primary role in the activation, magnitude, and velocity of slip \citep{passelegue2020initial, reches_fineberg} such ``that seismogenic faults can be activated by stress perturbations by all possible models of slip independently of frictional properties'' \citep{passelegue2020initial}.

The notion of strain energy as a primary factor controlling fault activity also bears on the formulation of the kinematic afterslip model.  To be clear, strain energy does not appear as a controlling parameter in the kinematic afterslip model in this paper. However, as soon as the afterslip period starts, there are no element-to-element interactions, and so the afterslip interval may be considered to be governed by local behavior.  Again, this is in contrast to mechanical afterslip models \citep[e.g.,][]{marone1991mechanics} where stress transfer during the afterslip epoch couples the fault system elements together and modulates their respective slip rates.  The fundamental geometric properties of strain energy  (rather than stress) transfer contribute to a rationale for this approximation of spatial independence.  Strain energy is given as $U = \frac{_1}{^2} \boldsymbol{\sigma} : \boldsymbol{\epsilon}$ where $\boldsymbol{\sigma}$ and $\boldsymbol{\epsilon}$ are the stress and strain tensors respectively.  This means that, in three dimensions, strain energy changes decay rapidly ($U{\sim}r^{-6}$) with distance, $r$ away from the source of the strain energy change due to the fact that strain decays as $r^{-3}$ \citep[e.g.,][]{love1927treatise, okada1992internal}.  This is relevant to the discussion of model locality because slip on an individual fault element may change the local strain energy significantly but will not provide significant coupling to other fault elements through strain energy coupling.  Thus, the spatially decoupled nature of the geometric moment-based afterslip model is more consistent with strain energy control of behavior than with stress-modulated activity. 

\section{Conclusions}
Afterslip following large earthquakes releases geometric moment in addition to that released coseismically.  Classical models for the location and rate of this release are predicated on the idea that afterslip is controlled by coseismic stress transfer and the rate-state frictional properties of fault zone materials in a way that predicts that co- and post-seismic slip should be largely spatially separated.  The geometric moment based model described here is independent of frictional properties and provides a means of describing the geodetic inferences of overlap of co- and post-seismic slip following the 2011 Tohoku-oki earthquake.

\clearpage
\bibliographystyle{agufull08}
\bibliography{references.bib}

\pagebreak
\section*{Acknowledgments}
The authors declare that they have no competing interests.  The authors acknowledge that they received no funding in support for this research.

\section*{Open research}
The software \citep{kinematic_afterslip_github} used for all calculations in the paper is preserved at DOI: 10.5281/zenodo.8173517, and developed openly at: \url{https://github.com/brendanjmeade/kinematic_afterslip}.

\pagebreak
\begin{table}
\caption{A comparison of stress/rate-and-state friction and kinematic afterslip models}
\label{tab:stress_vs_geometric_moment}
\centering
\begin{tabular}{ |l|c|c| } 
    \hline
      & rate and state & kinematic \\ 
    \hline
    stress-dependent & yes & no \\
    material properties dependent & yes & no \\
    afterslip outside coseismic region & yes & yes \\
    afterslip in coseismic region & no & yes \\
    segment coupling & stress transfer & fractional coseismic $m$ \\
    velocity decay with time & logarithmic & exponential \\
    \hline
\end{tabular}

\end{table}

%


\clearpage
\pagebreak

\pagebreak

\pagebreak

\pagebreak

\end{document}